\documentclass[manuscript,screen,acmsmall]{acmart}
\AtBeginDocument{%
  }

\setcopyright{acmlicensed}
\copyrightyear{2026}
\acmYear{2026}
\acmDOI{XXXXXXX.XXXXXXX}
\acmConference[CSCW '26]{Make sure to enter the correct
  conference title from your rights confirmation email}{June 03--05,
  2018}{Woodstock, NY}
\acmISBN{978-1-4503-XXXX-X/2018/06}

\begin{document}

\title{Interdependent Navigation and Pragmatic Disengagement: How Older Korean Immigrants Selectively Engage with Digital Technologies}

\author{Jeongone Seo}
\affiliation{%
  \institution{Rutgers University}
  \city{New Brunswick, NJ}
  \country{USA}}
\email{joh.seo@rutgers.edu}

\author{Tawfiq Ammari}
\affiliation{%
  \institution{Rutgers University}
  \city{New Brunswick, NJ}
  \country{USA}}
\email{tawfiq.ammari@rutgers.edu}

\renewcommand{\shortauthors}{Seo and Ammari}

\begin{abstract}
Older immigrant adults face unique barriers to digital participation, often framed as skill deficits. Through a community-based study with 22 older Korean immigrants in the greater New York area, we reframe these behaviors as active strategies. We identify pragmatic disengagement, where users selectively reject emotionally taxing or linguistically risky technologies, and interdependent navigation, where digital literacy operates as a distributed, relational resource rather than an individual skill. These practices reveal that non-use is often a culturally grounded form of "data refusal" shaped by values of dignity and family obligation. We contribute to CSCW by expanding theories of non-use beyond accessibility, offering design recommendations for "relational infrastructure" that supports dignity-preserving, collaborative engagement for aging immigrant populations.\end{abstract}




\received{20 February 2007}
\received[revised]{12 March 2009}
\received[accepted]{5 June 2009}

\maketitle

%% introduction_revised.tex
%% Revised Introduction section with changes highlighted in red.
%% Add \usepackage{xcolor} to the document preamble to render highlights.
%% \textcolor{red}{...} marks new or substantially revised text.
%% All original citations are preserved.

\section{Introduction}

The rapid expansion of digital technologies has created both opportunities and challenges for older adults. Central to this transformation are algorithmically mediated environments: recommendation systems that shape what content surfaces on platforms like YouTube, AI-powered voice assistants, automated financial and healthcare tools, and algorithmic curation systems that increasingly define the conditions of digital participation \cite{gillespie2014relevance}. From smartphones and social media platforms to these emerging artificial intelligence (AI) applications such as voice assistants \cite{10.1145/3373759}, recommendation algorithms \cite{10.1145/3677389.3702577}, and financial apps \cite{10.1145/3706598.3713427}, these tools are increasingly woven into daily life. However, many older adults find themselves on the margins of this digital transformation, with immigrant communities facing particularly acute challenges due to linguistic and cultural barriers \cite{10.1145/3613904.3642898,10.1145/3490555}.

Elderly immigrants frequently encounter multiple layers of disadvantage in digital adoption. Beyond the general digital divide—involving internet access, device ownership, and technical skills—they must navigate language barriers and cultural differences that compound difficulties in adopting new technologies \cite{chen2022use}. For older immigrants, algorithmically mediated systems introduce a further structural dimension to this exclusion: platforms trained predominantly on English-language data and Western behavioral norms may amplify emotionally charged content, misrecognize accented speech, or require identity verification procedures rooted in cultural assumptions that exclude users at precisely the moments of greatest vulnerability. Navigating these environments demands not merely technical skill but cultural and emotional discernment—a form of literacy that existing digital inclusion frameworks have been slow to recognize or support. These challenges are not uniformly distributed; they vary significantly across immigrant communities based on factors including language proximity to English, community institutional support, and geographic settlement patterns that affect access to culturally appropriate resources.

We focus on older Korean immigrants in the Greater New York City metropolitan area as a critical case for understanding these dynamics. This focus is motivated by three interconnected factors. First, the NYC metropolitan area hosts one of the largest and most established Korean immigrant communities in the United States, with over 200,000 Korean Americans residing in the region and a substantial elderly population \cite{jang2016risk}. This demographic concentration makes the community both significant in its own right and analytically valuable for understanding patterns that may inform research with other Asian immigrant groups. Second, the community maintains dense ethnic institutional infrastructure—including Korean-language temples, churches, cultural centers, and senior service organizations—that both facilitates community-based research and shapes the digital landscape elderly immigrants navigate. These institutions serve as crucial sites where technology adoption is negotiated, supported, or sometimes hindered. Third, prior research has documented high rates of social isolation among older Korean Americans \cite{jang2016risk}, yet their specific digital practices and the relationship between technology use and social connectedness remain critically understudied.

The Greater NYC area offers additional analytical advantages as a research site. As a global immigrant gateway with mature digital infrastructure, the region represents a ``most likely'' case where we would expect favorable conditions for digital inclusion—strong internet connectivity, proximity to tech support resources, and diverse language services. If elderly Korean immigrants experience significant barriers to digital participation even in this relatively resource-rich environment, it signals broader structural challenges that likely affect similar communities in less digitally developed regions. Moreover, the geographic concentration of Korean institutions in specific neighborhoods (such as Flushing, Queens and Fort Lee, New Jersey) \cite{NeilsbergKoreanPopNJCity,Wu2012AsianPopulation} creates distinct sociotechnical environments where community-specific norms around technology use develop and circulate, making visible the cultural dimensions of digital adoption that might be obscured in more dispersed populations.

In navigating this landscape, older Korean immigrants must contend not only with the practical challenges of smartphone use and platform interfaces but with the subtler terrain of algorithmic content curation, AI-assisted services, and automated systems whose design assumptions frequently exclude them linguistically and culturally. Understanding how this population selectively engages with—and strategically withdraws from—these algorithmically mediated environments offers insights not only for Korean immigrant communities but for aging immigrant populations more broadly, particularly those from non-English-dominant language backgrounds navigating similar structural exclusions. In this context, older Korean immigrants may experience heightened isolation if they are unable to engage effectively with digital platforms—yet they may also develop culturally-grounded strategies for selective engagement that existing frameworks overlook.

Recent studies underscore the triple-layered barriers older immigrants face when adopting digital technologies: unfamiliar technical features, limited linguistic support, and insufficient culturally responsive resources \cite{ekoh2023understanding,chen2022use}. During the COVID-19 pandemic, these disparities have only widened, revealing the urgent need for culturally tailored interventions that address both digital and AI literacy among immigrant elders \cite{nguyen2021digital}. 

Community-based approaches have shown promise in bridging these gaps. By involving local leaders and ethnic institutions, programs can be more relevant, trusted, and effective \cite{kuoppamaki2022enhancing}. Reflecting this perspective, the present study utilizes a community-based participatory research (CBPR) framework to explore how elderly Korean immigrants in Greater NYC engage with digital devices and services (including certain AI elements), and to identify pathways for more inclusive technology adoption. 

Existing literature documents several aspects of older adults' digital lives—including communication practices \cite{bucher2019algorithmic, Mittal_et_al,rohanifar2022kabootar}, misinformation exposure \cite{Mittal_et_al,le_at_al_24}, language-based exclusion \cite{shen_et_al_25}, and partial technology use. Yet these elements are typically examined in isolation, without attending to how algorithmic architectures, linguistic exclusion, and cultural values interact to produce the distinctive patterns of selective engagement and strategic withdrawal that characterize older immigrant communities navigating AI-mediated platforms. To better understand their interplay, we ask: 

\begin{quote}
\textbf{RQ1:} How do older immigrants navigate algorithmically mediated digital environments when faced with overlapping challenges of linguistic exclusion, cultural dissonance, and generational expectations? 

\textbf{RQ2:} In what ways do emotional and pragmatic forms of disengagement from digital platforms function as culturally-situated coping strategies among older adults, particularly within immigrant communities?  
\end{quote}

Our findings detail participants' daily digital practices, highlighting the multiple, intersecting forms of uncertainty they face and the coping mechanisms they employ. We draw on three interconnected bodies of theory that together form a coherent analytical framework: Uses and Gratifications Theory and Uncertainty Reduction Theory as a paired account of the motivational and uncertainty dimensions of digital engagement; scholarship on folk theories, selective appropriation, and episodic use as an account of the mechanisms through which partial domestication unfolds; and non-use and refusal scholarship as the conceptual basis for recasting strategic withdrawal as culturally-grounded agency. We then interpret these patterns through the lens of \textit{pragmatic disengagement}, arguing that what appears to be low digital literacy is often a strategic preservation of dignity and emotional well-being, enacted through \textit{interdependent navigation}—a mode of participation in which digital literacy is distributed across family, community, and commercial helpers rather than residing in the individual. We conclude with design and policy recommendations for creating the relational infrastructure necessary to support the interdependent nature of immigrant technology use.

%% related_work_revised.tex
%% Revised Related Work section with changes highlighted in red.
%% Add \usepackage{xcolor} to the document preamble to render highlights.
%% \textcolor{red}{...} marks new or substantially revised text.
%% All original citations are preserved.

\section{Related Work}
\label{sec:relatedword}

Understanding how people incorporate technology into their daily lives requires attending not only to individual decisions but also to the broader societal processes that shape them. As Bijker argues, values, skills, and goals are formed in local cultures, and we can therefore understand technological engagement by linking it to historical and sociological contexts \cite{bijker1997bicycles}. Domestication theory frames technology adoption as a process through which users incorporate new tools within the ``moral economy'' of everyday life—a set of values and social relations that shape how technologies are accepted, modified, or refused \cite{silverstone1992information, silverstone2006domesticating, haddon2007roger, bakardjieva2005internet}. People actively construct interaction patterns rather than acting as passive consumers \cite{jenkins2006convergence,bakardjieva2005internet}, reframing non-use not as failure but as active negotiation between individual agency and structural constraints.

This section reviews three interconnected bodies of scholarship that together constitute the analytical framework for this study. First, Uses and Gratifications Theory and Uncertainty Reduction Theory (\S\ref{UGTURT}) explain \textit{why} users selectively engage with digital technologies—what needs they seek to fulfill and how they manage ambiguity when linguistic and cultural uncertainty compound technical unpredictability. Second, research on folk theories, selective appropriation, and episodic use (\S\ref{sec:mentalmodel}) explains \textit{how} this selective engagement unfolds—through culturally mediated mental models, strategic feature adoption, and cyclical patterns of use and withdrawal. Third, scholarship on non-use and refusal (\S\ref{nonuse}) supplies the conceptual vocabulary for interpreting disengagement not as adoption failure but as culturally-grounded agency. Together, these three streams build toward our two core analytical constructs: \textit{pragmatic disengagement}—the selective refusal of emotionally or practically risky digital engagement—and \textit{interdependent navigation}—a distributed, relational mode of digital literacy in which the social network rather than the individual functions as the competent actor.

\subsection{Motivation and Uncertainty: What Users Seek from Digital Technologies}
\label{UGTURT}

Uses and Gratifications Theory (UGT) foregrounds user agency by emphasizing the personal needs and motivations behind technology use \cite{katz1973uses, rubin2009uses, lin2024unraveling}, distinguishing between anticipated and actual gratifications \cite{mclean2019hey}. Applied to AI systems including chatbots and voice assistants \cite{xie2023understanding, wei2024gratification}, UGT identifies three primary gratifications: \textit{social} (fulfilling relational needs through connection and relationship-building) \cite{xie2023understanding, IFINEDO2016192}; \textit{utilitarian} (increasing efficiency by offloading routine tasks) \cite{ammari_et_al_19, BRACHTEN2021102375}; and \textit{hedonic} (deriving enjoyment from playful interactions or immersive experiences) \cite{oewel2023voice, jo2022continuance, li2015modeling}.
Uncertainty Reduction Theory (URT) complements UGT by explaining how users manage ambiguity through interactive strategies (direct engagement) and passive strategies (observing others) \cite{chang2024uncertainty, kramer1999motivation}. Where UGT illuminates what users seek from digital tools, URT illuminates how they navigate unpredictability when those tools behave ambiguously—a pairing that is especially productive for analyzing older immigrants, for whom linguistic and cultural distance layer additional uncertainty onto the technical ambiguity all users face.

\paragraph{Social Gratification}
Social technologies fulfill relational needs supporting healthy aging \cite{diener2009introduction}, particularly addressing loneliness in care settings \cite{bar2021fostering, creighton2016prevalence}. Online engagement can reduce loneliness and anxiety \cite{sum2008internet, heo2015internet, bell2013examining}, offering connections, advice, and support \cite{antonucci2017social, chopik2016benefits, czaja2018improving, mitzner2019technology}. However, older immigrants express concerns about privacy, trust, and security that influence adoption decisions \cite{bixter2019understanding}.

\paragraph{Utilitarian Gratification}
Older immigrants often seek healthcare support through technology. AI health assistants can reduce caregiving burdens with stable infrastructure and family assistance \cite{bults2024acceptance}, though access to virtual healthcare varies by immigration status, language skills, and length of stay \cite{brual2023virtual}. Adoption is shaped more by socioeconomic status, language barriers, and cultural misalignment than ethnicity \cite{hawkins2022barriers}, while systems often lack culturally responsive design \cite{chen2022use}.
Older adults face heightened misinformation exposure not from individual deficits but because algorithms amplify emotionally charged content \cite{lazer2018science}, leading to fatigue and intentional disengagement \cite{rohanifar2022kabootar, joy_et_al_25, bucher2019algorithmic}. McKenna et al. show older adults prefer interpersonal validation over algorithmic content \cite{mckenna2025chinese}, while Liaqat et al. highlight confusion about privacy and control \cite{Liaqat_et_al_21}. Withdrawing from digital platforms often represents rational decisions to preserve emotional well-being.

\paragraph{Hedonic Gratification}
Multimodal technologies offered enjoyment and cultural continuity during COVID lockdowns \cite{richards_et_al_21}. Blogging and storytelling helped older adults explore identity \cite{brewer_et_al_16}, while everyday tasks like emailing offered pleasure when perceived as valuable \cite{zhang2012older}. Photo-sharing helped construct meaning and personal connections \cite{waycott_et_al_13, zhang2023older}, though emotional safety, complexity, and cultural fit shaped which activities were pursued \cite{harris2022older, chen2022use}.

Across these three gratification types, a consistent pattern emerges that will structure our analysis: digital engagement is embraced when it delivers clear relational or practical value under conditions of manageable uncertainty, and abandoned when linguistic barriers, algorithmic manipulation, or cultural misalignment tip the cost-benefit calculus toward disengagement. This motivational framework—the \textit{why} of selective engagement—sets up the following section's account of the \textit{how}.

\subsection{Mechanisms of Selective Engagement: Folk Theories, Domestication, and Temporal Rhythms}
\label{sec:mentalmodel}

If UGT and URT explain the motivational calculus behind selective engagement, the frameworks reviewed here explain the mechanisms through which it is enacted. Users do not simply decide to adopt or reject technologies in a single moment; they construct culturally shaped mental models of opaque systems, strategically appropriate specific features while ignoring others, and cyclically engage and withdraw to manage their well-being over time. For older immigrants navigating AI-mediated platforms, these mechanisms are particularly consequential: the distance between users' cultural and linguistic frameworks and systems' design assumptions shapes not only which features feel accessible, but which feel safe.

\subsubsection{Culturally-Mediated Folk Theories}

Folk theories—intuitive mental models users construct to interpret opaque system behaviors—are shaped as much by cultural context as by direct experience \cite{johnson1989mental, gelman2011concepts}. Culture mediates how users form expectations, calibrate trust, and interact with intelligent technologies \cite{devito2017algorithms, eslami2016first, karizat_et_al_21}, often through tacit reasoning they cannot consciously articulate \cite{devito2018people, rader2015understanding}. These culturally grounded theories manifest in metaphorical reasoning: users describe AI as ``overwhelming'' or ``attacking'' \cite{druce2021brittle}, while content creators construct ``algorithmic personas'' casting platforms as agents or gatekeepers \cite{cotter2019playing}—metaphors that both scaffold understanding and expose its limits \cite{desai_et_al_metaphor_23}.

Yet folk theories frequently diverge from technical reality. AI systems lack the social grounding and contextual nuance users assume \cite{riedl2019human}, missing sarcasm \cite{jena2020c} or misrecognizing accented speech \cite{pal2019user, mittal_et_al_25}—failures that are not culturally neutral, as they fall disproportionately on users whose linguistic and social norms diverge from those encoded during training. Such brittleness erodes trust \cite{druce2021brittle}, which users continuously revise as systems confirm or violate their expectations \cite{bucinca2021trust, chong2022human}. Cultivating accurate, culturally informed mental models is therefore essential for effective human-AI collaboration \cite{nourani2021anchoring, desai2024mental}.

\subsubsection{Selective Use as Strategic Incorporation}
Selective use represents ``partial domestication'' where users engage with specific affordances while rejecting others. Salovaara et al. describe this ``appropriation work'' \cite{balka2006making, salovaara2011everyday} where users repurpose technology (e.g., using smartphones but disabling social notifications) to prevent technology from becoming intrusive. This aligns with S{\o}rensen's view that domestication involves users disciplining artifacts to serve the household's moral economy \cite{sorensen2006domestication, silverstone1992information}.

\subsubsection{Episodic Use and Self-care}
While traditional domestication implies stabilization over time, episodic use highlights temporal cycles of engagement and abandonment. Rather than viewing gaps as adoption failures, recent scholarship frames episodic use as ``self-care'' \cite{gorm2019episodic}. Users ``pause'' domestication when constant monitoring produces anxiety—the ``burden of tracking'' \cite{epstein_2020}. This regulatory mechanism protects well-being, adding cyclical dimensions to the Incorporation phase where users may domesticate devices for specific goals, release them during non-use, and re-domesticate later when needed. Gorm and Shklovski extend this framework, showing that episodic use among older adults constitutes a deliberate practice of care rather than technological abandonment—a framing that directly informs our interpretation of participants who disengage from YouTube during periods of political turmoil and re-engage when content feels emotionally safe and culturally resonant \cite{gorm2019episodic}.

Together, folk theories, selective appropriation, and episodic use constitute the mechanisms through which users achieve partial domestication under conditions of cultural and linguistic uncertainty. They illuminate how users develop protective heuristics, limit engagement to familiar features, and cyclically withdraw to preserve well-being—patterns that will recur directly in our participants' practices with KakaoTalk, YouTube, and emerging AI tools. What these mechanisms produce, however, can look from the outside like non-use or failure. The following section provides the conceptual tools to reinterpret such outcomes as culturally-grounded agency.

\subsection{Selective Use, Non-Use, and Refusal}
\label{nonuse}

The mechanisms described above frequently result in patterns that adoption-centric frameworks misread as failure. This section reviews scholarship that reframes deliberate disengagement as meaningful, agentic behavior—establishing the conceptual basis for our constructs of pragmatic disengagement and interdependent navigation.

HCI scholarship increasingly recognizes that understanding refusal, limiting, or abandonment is as important as studying adoption. Baumer et al. introduced categories like ``lagging resistance'' and ``symbolic non-use'' to explain deliberate disengagement \cite{baumer_et_al_13}, framing non-use as meaningful behavior and introducing terms like ``resisters'' and ``liminal users'' \cite{baumer_et_al_14}. Densmore examined when technologies should fail \cite{densmore2012}, while Barocas et al. frame refusal as ethical responsibility for practitioners \cite{barocas2020}.

Crucially, however, not all non-use is the same. McGranahan \cite{mcgranahan2016theorizing} draws an important distinction between resistance and refusal: while resistance presumes a hierarchical relationship in which a subordinate pushes back against a superior power, refusal ``works differently by professing a relationship between equals'' and operates by ``redirecting levels of engagement'' rather than simply opposing them \cite{mcgranahan2016theorizing}. Refusal, in this framework, is generative—``the ending of one thing is often the generation of something new''—and affiliative, producing new forms of community and shared meaning among those who refuse together \cite{mcgranahan2016theorizing}. We argue that participants' withdrawal from financially risky platforms, algorithmically manipulated feeds, or cognitively overwhelming interfaces represents refusal as defined by McGranahan—not a failure to engage, but an active reorientation of engagement on their own moral and relational terms.

Building on DiSalvo's conception of design as space for contestation \cite{disalvo2015adversarial}, Zong and Matias articulate ``data refusal from below'' centering resistance to algorithmic systems \cite{Zong_Matias_24}, while Garcia et al. position refusal as feminist practice \cite{garciaetal2020}. Light and Akama argue that designing with communities navigating structural barriers requires emotional attunement, as digital practices are shaped by shared histories and affective ties \cite{light_akama_2012}.

Empirical work documents refusal across contexts. Sampson et al. found queer users negotiate representation dynamics with targeted advertising \cite{sampsondenae23}, Johnson et al. characterized system abandonment \cite{johnson1989mental}, and Cha and Wong examined AI non-use in professional settings \cite{chatetal25}. In participatory design, Robinson et al. explored participation through refusing \cite{robinsonetal2020}, while Dourish et al. highlighted affective demands on repeatedly researched communities \cite{dourishetal2020}.

Critically, disengagement is relational and contextual. Light and Akama show non-use emerges from social circumstances \cite{light_akama_2012}. Brewer et al. found older adults avoid online discussions to manage tensions, preferring phone calls \cite{brewer_et_al_21}. Sheehan and Le Dantec describe ``functional minimalism''—selecting only manageable tools \cite{sheehan_23}. Zhao et al. show older migrants rely on family or analog tools to maintain control \cite{zhao_et_al_23}.

Language and cultural positioning profoundly shape engagement. Older users depend on family for translation and interpretation \cite{wong-villacres_et_al_19, yuan_et_al_24}, while English-centric medical portals create barriers forcing reliance on caregivers \cite{le_et_al_24}. Community organizations bridge gaps left by technologies designed without immigrant communities \cite{villanueva_et_al_23}. Voice assistants that misinterpret accented speech drive abandonment \cite{mittal_et_al_25}, and older migrants often distrust online technologies, preferring familiar platforms and relying on family or analog tools \cite{chi_ict_24, hsiao_et_al_23}. These patterns reveal language-based non-use reflects systemic design exclusions rather than skill deficits.

Beyond language-based exclusion, we observe a culturally grounded form of what we term algorithmic sovereignty—the deliberate refusal of platform logics on the platform's terms, paired with the active generation of alternative informational practices rooted in community trust and peer-vetted knowledge. Drawing on McGranahan, this is not resistance against a dominant system but a refusal that ``establishes a relationship of equals'' \cite{mcgranahan2018refusal}: participants who bypass recommendation algorithms and instead circulate links through KakaoTalk, discuss content collectively at senior centers, or apply community-developed heuristics for evaluating trustworthy sources are asserting their own conditions of engagement. This refusal is generative: the rejection of algorithmic curation produces an alternative knowledge infrastructure embedded in social relationships and cultural values. 

This body of scholarship—spanning non-use typologies, the resistance/refusal distinction, data refusal frameworks, and relational accounts of disengagement—provides the conceptual foundation for the two constructs our empirical analysis develops. \textit{Pragmatic disengagement} names the culturally situated refusal of emotionally or practically risky digital engagement documented in our data: neither passive withdrawal nor simple resistance, but a generative act that produces alternative modes of participation rooted in dignity, relational harmony, and Korean communal values. \textit{Interdependent navigation} names the distributed mode of digital literacy that sustains this selective engagement: a pattern in which the family unit or social network, rather than the individual, functions as the digitally competent actor—extending existing accounts of intergenerational mediation \cite{wong-villacres_et_al_19, yuan_et_al_24} by foregrounding the emotional costs and dignity stakes of help-seeking. Together, these two constructs synthesize the motivational account of UGT and URT, the mechanistic account of domestication and folk theory scholarship, and the normative reframing of non-use and refusal scholarship, showing how all three streams converge to illuminate how older Korean immigrants participate in digital life on their own cultural terms.

This paper explores how selective engagement and strategic non-use unfold within Korean immigrant older adult communities, where linguistic exclusion, emotional fatigue, and relational dependencies converge to produce two core practices: pragmatic disengagement—a culturally situated refusal of emotionally or practically risky digital engagement—and interdependent navigation—a mode of participation in which digital literacy is distributed across family, community, and commercial helpers rather than residing in the individual.

\section{Methods}
This section outlines the methodological approach used to examine digital and AI literacy among older Korean immigrants. Guided by community-based participatory research (CBPR), the study was designed in close collaboration with community partners and grounded in principles of cultural responsiveness and mutual trust. We describe our research design, participant recruitment strategies, data collection procedures, and thematic analysis process. Together, these components enabled an in-depth understanding of participants’ everyday digital practices, challenges, and coping strategies.

\subsection{Researcher Positionality and Community Engagement}
The first author is a Korean immigrant who resides in the U.S. for and is a Licensed Master Social Worker (LMSW) in the Northeast. This dual identity as a community insider and a trained researcher shaped the study's engagement. The second author, a cisgender male, investigates the dual nature of social technologies in their capacity to empower marginalized populations while simultaneously creating vulnerabilities related to privacy, personal autonomy, and identity. 

In this paper, our community partners include leaders and staff at two local organizations in the Northeast of the United States: a Buddhist temple (abbot and lay volunteers) and an adult day-care center (director and care workers). Following CBPR principles, these partners were engaged as co-researchers across the study cycle: they helped refine the focus and recruitment/rapport-building activities, co-developed the interview guide, and participated in member-checking of preliminary themes \cite{israel1998review,wallerstein2017community}. 

The partnership with the Buddhist temple in the Northeast region of the US grew from a pre-existing relationship of trust; the first author frequently visited the temple for personal meditation. The abbot, a respected master of Buddhist meditation, viewed the research as socially necessary because many temple members had expressed frustration with digital technologies and a desire for guidance. Given his role in connecting community members with education al and social resources, he saw this study as an opportunity to address an unmet need.  

Entry into the adult day-care center in the Northeastern city occurred organically. While the author was distributing recruitment flyers outside a Korean grocery store, the director of a local day-care center received one. They subsequently invited the first author to conduct the study at their facility, framing the interviews as a special program for the residents. This framing helped normalize participation by integrating the research into the center’s existing enrichment programming, reducing any stigma associated with discussing technological difficulties. 

In this paper, our community partners include leaders and staff at two local organizations in the Northeast of the United States: a Buddhist temple (abbot and lay volunteers) and an adult day-care center (director and care workers). Following CBPR principles, these partners were engaged as co-researchers across the study cycle: they helped refine the focus and recruitment/rapport-building activities, co-developed the interview guide, and participated in member-checking of preliminary themes.

\subsection{Research Design and Participant Engagement}
This study employed a qualitative research design grounded in the principles of community-based participatory research (CBPR). Drawing from established CBPR frameworks in both HCI and public health \cite{wallerstein2017community,zhao_et_al_23}, this approach enabled culturally resonant inquiry and strengthened mutual trust between the research team and community collaborators. In line with best practices, participants were treated not only as informants but as co-researchers who contributed to the interpretation of cultural meaning and research direction.

Partners from the Buddhist temple and the adult day-care center were engaged as co-researchers throughout the process. During the formative stage, the first author met twice in person with the temple's abbot to collaboratively decide on: (a) the mode of inquiry (survey vs. one-to-one interviews), (b) participant recruitment routes (abbot's personal introductions vs. open sign-ups), and (c) rapport-building activities prior to data collection. The partnership agreed to pursue semi-structured one-to-one interviews, combine personal introductions with voluntary sign-ups, and host three temple-based tea-tasting gatherings of --- Pu-erh in Korean. The choice of \textit{Pu-erh} was culturally significant, as this prized, aged fermented Chinese tea is central to the temple’s tea meditation practice \cite{tearepertoire2024tteokcha,pathofcha2024mindfulness,cspuerh2024ceremonies}. This helped the researcher become a familiar presence in the community. In parallel, the director and care workers at the adult day-care center advised on the appropriate timing and framing for inviting older adults within their daily program. These processes exemplify the CBPR standards of equitable power-sharing and co-learning \cite{wallerstein2017community,israel1998review}. All study procedures were approved by the university's Institutional Review Board (Protocol Pro2024001102).

\subsection{Participant Recruitment}
Between November and December 2024, 29 Korean immigrant older adults (aged 65-90s) were recruited through purposive and snowball sampling across two community sites in the Greater New York City metropolitan area: a Korean Buddhist temple and a senior day-care center. Participants resided in New Jersey, New York, and Pennsylvania. These sites were selected due to the first author's established community ties and the dense concentration of Korean immigrant institutions in this region, which facilitated trust-based recruitment. 

Community leaders—trusted figures within these institutions—played an instrumental role in facilitating introductions and validating the research purpose to potential participants \cite{zhao_et_al_23}. At the temple, the abbot announced the study during services and used tea-tasting gatherings to facilitate informal introductions with the researcher, reducing refusal anxiety. At the day-care center, staff framed participation as a "smartphone use conversation/coaching" session and integrated interview slots into the daily program schedule. All participation was voluntary with no consequences for declining, and each participant received a \$25 incentive in cash or grocery store gift cards. 

In total, 22 interviews were successfully recorded and analyzed (conducted either in person or by phone). Seven additional telephone interviews \cite{novick2008there} were excluded from formal coding due to a recording app malfunction that deleted the audio files, though expanded field notes were retained to provide contextual understanding \cite{halcomb2006verbatim}.

Table \ref{tab:participants} summarizes the demographic profiles and digital usage characteristics of the 22 participants whose interviews were included in the analysis. 

\begin{table}[h]
\scriptsize
\centering
\caption{Summary of Participant Demographics, Technology Practices, and Use Contexts. This table provides an overview of the 22 older Korean immigrant participants in our study, including their approximate age group, gender, and their recruitment site. The "Tech Use" column  provides contextual detail on how and why these technologies are incorporated into their daily lives. Patterns reflect a wide range of digital engagement, from practical communication and spiritual media consumption to intentional resistance or selective non-use. Note: KCC stands for Korean Community Center}\begin{tabular}{|c|c|c|c|p{2.5cm}|p{10cm}|}
\hline
\textbf{ID} & \textbf{Age Group} & \textbf{Gender} & \textbf{Recruitment Site}  & \textbf{Tech Use} \\
\hline
OA01 & Late 60s & F & Buddhist Community, NJ  & Uses financial apps; interested in cultural/spiritual heritage \\
\hline
OA02 & Early 70s & F & Buddhist Community, NJ  & Listens to Buddhist content, self-directed learning, studied Excel \\
\hline
OA03 & Early 90s & M & Buddhist Community, NJ &  Retired civil servant; talks about intergenerational conflicts \\
\hline
OA04 & Early 70s & F & Senior Daycare Center & Experienced KakaoTalk mishaps; fatigued by repetitive learning \\
\hline
OA05 & Mid 60s & F &  Buddhist Community, NJ & Proactive learner; shares health info via social media \\
\hline
OA06 & Early 70s & F & Senior Daycare Center  & Difficulty resolving issues independently \\
\hline
OA07 & Late 60s & F & Buddhist Community, NJ & Strong digital security awareness; fluent in Korean and English \\
\hline
OA08 & Early 70s & F & Buddhist Community, NJ & Focuses on health content; shares with family \\
\hline
OA09 & Late 60s & F & Buddhist Community, NJ  & Distrusts online transactions; struggles with English interfaces \\
\hline
OA10 & Early 60s & F & Buddhist Community, NJ  & Prefers intuitive features like messaging and calling \\
\hline
OA11 & Mid 70s & F & Senior Daycare Center  & Uses recommendations; limits tech engagement \\
\hline
OA12 & Mid 70s & F & Senior Daycare Center & Stopped using apps after a group chat mishap \\
\hline
OA13 & Early 70s & F & Senior Daycare Center  & Files detailed technical complaints and requests \\
\hline
OA14 & Mid 60s & F & Buddhist Community, NJ & Avoids digital tools due to strong security concerns \\
\hline
OA15 & Mid 70s & F & Senior Daycare Center & Learns tech via feedback/help from family \\
\hline
OA16 & Late 60s & F &  Buddhist Community, NJ  & Actively searches health/cooking content \\
\hline
OA17 & Early 70s & F & Senior Daycare Center & Engages in cultural/physical activities for cognitive health \\
\hline
OA18 & Mid 60s & F & Buddhist Community, NJ  & Frequently uses digital media for music/video \\
\hline
OA19 & 70s & F & Buddhist Community, NJ  & Digital use focused on Buddhist content \\
\hline
OA20 & Early 80s & M & Senior Daycare Center  & Limited digital tool use overall \\
\hline
OA21 & Late 70s & F & Senior Daycare Center  & Self-assured but needs repeated instruction \\
\hline
OA22 & Late 70s & F & Senior Daycare Center  & Uses tech to reduce generational gap \\
\hline
\end{tabular}
\label{tab:participants}
\end{table}

\subsection{Data Collection}
Semi-structured interviews were conducted in Korean by the first author, a bilingual social worker and doctoral researcher with established ties to the local Korean immigrant community. Each interview lasted approximately 45–60 minutes and was conducted in locations familiar to participants—such as homes, temples, senior centers, and grocery store seating areas—or via telephone when preferred. All interviews were audio-recorded with participant consent, except in cases of technical malfunction, where detailed field notes were taken instead. 

The interview guide focused on five main areas: 
\begin{enumerate}
    \item Comfort and frequency of technology use (e.g., smartphones, computers)
    \item Experiences with or perceptions of AI tools (e.g., voice assistants, health apps)
    \item Barriers to learning and using new technologies
    \item Sources of technical help (e.g., family, caregivers, community members)
    \item Attitudes toward emerging technologies in health and communication contexts
\end{enumerate}

This structure allowed for conversational flexibility while ensuring consistency across cases. Participants often spoke at length, offering both descriptive and reflective narratives. The linguistic match between researcher and participants enhanced rapport and reduced barriers to expressing concerns and frustrations \cite{icma2020digitaldivide}. 

All participants gave verbal and written consent prior to participation. Confidentiality was strictly maintained through anonymization and encrypted data storage. The researcher’s dual identity—as both insider (Korean, immigrant, community member) and outsider (doctoral researcher)—was explicitly acknowledged.

\subsection{Data Analysis}
Interview data were analyzed using inductive thematic analysis \cite{braun_one_2021}. While open coding proceeded without predetermined categories, concepts from the literature — including uses and gratifications, layered uncertainty, and the distinction between refusal and resistance — served as sensitizing frameworks during the later stages of theme development and cross-case comparison. This hybrid approach allowed empirical patterns to emerge from the data while enabling us to situate findings within relevant theoretical discourse. The coding was conducted in Korean by the first author, ensuring the preservation of linguistic nuance and context. The analytic process followed six stages: (1) data familiarization, (2) open coding, (3) code consolidation, (4) theme development, (5) cross-case comparison, and (6) cultural validation. 

Themes were developed iteratively, with close attention paid to culturally embedded language and recurring narrative patterns. To enhance validity, preliminary findings were reviewed by three community advisors—a senior center volunteer, a Korean-American caregiver, and a bilingual pastor serving elderly congregants—who helped interpret findings in light of community-specific values. A similar process was adopted by Zhao et al. \cite{zhao_et_al_23}. 

All excerpts cited in the Findings section were translated into English by the research team, with efforts made to retain the emotional tone, context, and voice of each participant. 

Reflexive memos were used to track emerging interpretations and minimize bias. In line with participatory research protocols, community members had the opportunity to review and comment on emergent findings prior to dissemination\cite{wallerstein2017community,zhao_et_al_23}. Between coding rounds, the research team met to review and refine the coding schema. Subsequent rounds of coding focused on identifying thematic patterns, which were grouped into broader categories after reaching consensus \cite{mcdonald_et_al_2019}. At this stage, theoretical frameworks introduced in \S\ref{sec:relatedword}, particularly episodic use \cite{gorm2019episodic}, pragmatic disengagement, and refusal \cite{mcgranahan2016theorizing}, were brought into dialogue with the empirical themes, not to impose predetermined categories but to sharpen the analytical language through which we describe and interpret participant experiences. 

\section{Findings}
This section presents the lived experiences of older Korean immigrants as they navigate digital technologies in their daily lives. Drawing from interviews, we highlight patterns of everyday use, barriers encountered, and the social and emotional strategies participants employ to adapt to or disengage from digital platforms. These findings are organized around three thematic areas: daily digital practices (\S\ref{findings:everyday}), barriers and coping mechanisms (\S\ref{findings:barriers}), and evolving perspectives on media and emerging technologies (\S\ref{findings:evolving}).

\subsection{Everyday Digital Use Patterns}
\label{findings:everyday}
We begin by examining the day-to-day ways in which participants use smartphones and digital media. Their engagement is not characterized by an effort to master every available feature or application, but rather reflects a highly selective and pragmatic process of adopting a small suite of tools that serve immediate and meaningful needs. This selective adoption reflects a clear hierarchy of values where technology is prized primarily for its ability to maintain cherished social ties, access culturally resonant information, and manage essential communication through familiar, comfortable, and trusted tools, with other digital possibilities often being deemed irrelevant or not worth the effort.

As we will show, participants' choices align closely with the gratification categories identified by Uses and Gratifications Theory (\S\ref{UGTURT}): KakaoTalk fulfills social needs, YouTube serves both enjoyment and practical information-seeking, and visual media exchanges extend social connection across transnational distances. At the same time, participants engage in selective appropriation (\S\ref{sec:mentalmodel})—adopting specific features while ignoring others—revealing how domestication unfolds as ongoing negotiation rather than wholesale acceptance.

\subsubsection{A Central Social Hub and a Lifeline: KakaoTalk for Family and Community Ties Locally and Internationally} 

For many older Korean immigrants, the smartphone functions as a lifeline for maintaining transnational family ties, managing daily communication, and accessing culturally relevant information. However, rather than exploring a broad ecosystem of apps, participants rely on a small set of familiar tools, with KakaoTalk serving as the undisputed hub of this activity. This mobile messenger application, dominant in South Korea and ubiquitous within the Korean diaspora, functions as the primary artery for communication. As multiple participants stated, their core uses are simple and focused: "Phone calls, text messages, KakaoTalk" (OA03); "I use YouTube, KakaoTalk, and text messages" (OA22).

The preference for KakaoTalk over traditional SMS is rooted in practical advantages: it seamlessly supports the Korean language, facilitates easy exchange of photos and videos, and, most critically for an immigrant community, offers free international voice and video calls. One participant explained its fundamental appeal with simple clarity: "It's convenient because I can call family without any charge" (OA04). For a community with deep ties to their country of origin, this feature is not a mere convenience but an essential tool for maintaining family cohesion. This pattern exemplifies the social gratification dimension (\S\ref{UGTURT}), where technology fulfills relational needs through connection and relationship-building.

These digital interactions are not merely functional; they are laden with emotion and serve to bridge vast physical distances, creating a palpable sense of presence and connection. Participants regularly described sending and receiving photos of grandchildren, shared meals, or scenes from their daily lives as a way to participate in milestones and offer comfort (OA15). One participant detailed a common practice that illustrates how these exchanges blend emotional connection with practical problem-solving: "I send photos of family moments, meals, or even household problems—such as a broken faucet—to receive help or simply feel connected. For example, I sent a photo of a broken faucet to my daughter, who responded with a how-to video" (OA15). This interaction transforms digital exchange into a moment of remote, collaborative caregiving and problem-solving within transnational family networks.

While central to their lives, participants' use of KakaoTalk often remained functionally minimalist—a clear instance of selective appropriation (\S\ref{sec:mentalmodel}), where users engage with specific affordances while rejecting others. Advanced features like paid emoticons, calendar integrations, or complex voice-to-text functions were frequently ignored, viewed as confusing, or deemed unnecessary. One participant described their frustration with the voice-to-text feature, noting a common issue for bilingual users: "English words get oddly converted" (OA02). This illustrates how even minor usability barriers, particularly those related to language, can effectively wall off certain features, reinforcing a pattern of using only the most basic and reliable functions. This focused use reveals a broader logic: technology is warmly embraced when it fulfills clear and immediate social and familial needs, while less familiar or more complex functions are either delegated to others or deliberately and pragmatically avoided—a practice that could be called selective digital participation, guided by cultural familiarity, emotional value, and technical comfort.

\subsubsection{YouTube as Selective Information Curation and Community Knowledge-Sharing} 
YouTube plays a central role in the digital lives of older Korean immigrants, serving multiple needs simultaneously (\S\ref{UGTURT}). It functions as far more than entertainment—serving as an essential, on-demand information commons, a personalized classroom, and a culturally specific media outlet that offers enjoyment and cultural continuity. For many, it provides an accessible, Korean-language platform for a vast range of needs, from staying current with news and politics in Korea to learning new skills and satisfying personal curiosities. As participants described, their viewing habits were diverse and purposeful: "I watch a variety of things. Health programs, golf programs, cooking shows" (OA16); "I mostly listen to [religious] sermons on YouTube... and watch the [Buddhist television channel] Live broadcast" (OA19).

Crucially, YouTube functions not just as an individual learning tool, but as a shared \textit{digital community bulletin board}. The culturally relevant information gathered from the platform—be it political debates, health tips, or religious sermons—often forms the basis for later discussions with family members, friends at the day-care center, or fellow members of their religious communities. One participant noted that staying informed on Korean news was a vital way to remain connected to ongoing community discourse, both locally and with relatives in Korea (OA10). This social dimension of YouTube use extends the social gratification function beyond direct messaging platforms.

However, participants were not passive consumers of this information. Many expressed frustration with the platform's recommendation algorithm, which they felt often surfaced sensational, politically biased, or simply irrelevant content. In response, they developed sophisticated strategies to manage the platform, revealing the development of accurate mental models about algorithmic systems (\S\ref{sec:mentalmodel}). A common tactic was bypassing the algorithm-driven recommendation feed entirely and using manual search functions to find specific, trustworthy content. As one participant explained: "When too many suggestions appear, it gets overwhelming... so I prefer to search myself. Since videos with high view counts are shown first, you might only see sensational content. That's why I deliberately try different searches" (OA02).

Another participant elaborated on their personal filtering process, based on established trust and personal values rather than algorithmic suggestion: "I search for what I need... I mostly watch people I've seen before. For cooking, there are a few people I subscribe to. Even if a video is famous, if it doesn't align with my values, like using too much sugar, I don't watch it. I have to agree with it" (OA16). This self-regulated use reveals clear awareness of algorithmic influence and a cautious, critical approach to media consumption—exemplifying selective appropriation strategies (\S\ref{sec:mentalmodel}), where users repurpose technology to prevent it from becoming intrusive.

Participants often disengaged from videos that appeared misleading, especially around sensitive topics like health or politics, prioritizing content that was practical, culturally resonant, and delivered in Korean while steering clear of material that felt emotionally taxing or unreliable. One participant captured this critical mindset: "Especially with health information, there are so many things that seem baseless... it just makes my head more complicated, so after watching for a bit, I decide it's better not to watch. If I think, 'this isn't quite right,' I just turn it off" (OA06).

This pragmatic stance toward media engagement—selectively engaging, co-interpreting, and disengaging when uncertainty or affective cost grows—was particularly evident during the brief martial law declaration in South Korea on December 3, 2024. Participants turned to a small set of Korean-language YouTube channels pre-vetted by family members and friends for rapid interpretation, receiving links via KakaoTalk followed by in-person discussion at day-care centers and temples. YouTube braided with KakaoTalk operated as an event-driven rumor-and-interpretation infrastructure that moved through tightly knit community ties, with participants initially repeating rumors of a "second declaration" (fieldnotes, Dec 2024). However, pushback from peers and family subsequently tightened shared heuristics—manual search rather than home feed browsing, sticking to known channels—practices participants described as "sticking to what we know" to avoid clickbait and emotional overload while aligning viewing with group norms and reducing conflict.

\subsubsection{Social Networking and Photo/Video Exchange} 
Visual media—especially photos and short videos—play a central role in how older Korean immigrants maintain emotional intimacy and social presence with loved ones, extending social gratification (\S\ref{UGTURT}) across transnational distances. While not always engaged in broader social networking platforms, participants actively used tools like KakaoTalk for private exchanges within family groups or close-knit circles. These interactions were not primarily about social performance or public updates, but about sustaining everyday bonds in diasporic contexts.

Most participants described regularly sending and receiving photos of grandchildren, meals, or household events through KakaoTalk. These images served multiple purposes: they allowed participants to celebrate family milestones, share updates, and solicit advice on daily matters. For example, one participant mentioned sending a photo of a broken faucet to a daughter, who responded with a how-to video, effectively transforming the exchange into a remote caregiving moment. These interactions reinforced interdependence across distance, blending emotional reassurance with practical problem-solving.

Despite the centrality of these exchanges, engagement varied in depth and frequency—reflecting patterns of partial adoption (\S\ref{sec:mentalmodel}). Some participants were highly active in group chats, exchanging daily greetings, short videos, or celebratory stickers. Others maintained a more utilitarian approach—only sending images when necessary or when prompted by family. Advanced features like animated GIFs, filters, or camera effects were often seen as unnecessary or confusing, and were largely avoided.

Underlying these patterns is a clear logic of pragmatic selectivity. Visual exchanges are embraced when they serve tangible functions—connecting families, conveying emotions, or resolving issues. Yet more playful or complex features are often bypassed, reflecting an emphasis on communication over novelty. As one participant noted, it is not about keeping up with trends, but about "sharing what matters, when it matters."

Taken together, these practices demonstrate how older immigrants adapt digital media to fit their social and emotional priorities. In contexts marked by separation, aging, and language barriers, visual communication becomes a powerful tool for presence, support, and continuity—used not to explore the full range of digital possibilities, but to reinforce the ties that matter most.

\subsection{Barriers and Coping Mechanisms: The Centrality of Interdependent Navigation}
\label{findings:barriers}
Despite their regular and often sophisticated use of a select set of digital tools, participants faced persistent and formidable barriers that limited their ability to engage more fully with the digital world. These hurdles included systemic language gaps that went far beyond simple translation, the overwhelming complexity of modern user interfaces, and the relentless pace of technological change. 

Building on Uncertainty Reduction Theory (\S\ref{UGTURT}), participants managed multiple, intersecting forms of ambiguity: technical uncertainty (system interactions), linguistic uncertainty (English-dominant interfaces), relational uncertainty (social costs of help-seeking), and cultural/epistemic uncertainty (credibility and values alignment). Their primary coping mechanism for overcoming these challenges was interdependent navigation—a fundamentally collaborative mode of digital participation that is heavily scaffolded by social networks and positions digital literacy as a distributed, rather than individual, competency (\S\ref{nonuse}). This section outlines how participants adapted to these barriers through family support, community knowledge-sharing, or sometimes intentional non-use, while exploring the significant "relational costs" that often accompany this reliance on others.

\subsubsection{"It's a Headache": The Emotional and Cognitive Burden of Learning New Digital Skills}
While participants expressed comfort and proficiency with the tools they used daily, the prospect of learning a new feature, installing a new application, or troubleshooting an unfamiliar problem was often met with profound dread and exhaustion. The difficulty lay not merely in the technical steps but in the cumulative cognitive and emotional demands of adapting to constantly changing and unforgiving digital environments. These experiences exemplify challenges in uncertainty reduction (\S\ref{UGTURT}), where users struggle to manage ambiguity.

Installing a new app, for instance, was described as a gauntlet of English-language prompts, account creation forms, password requirements, and security code verifications. These layered requirements, presented in a non-native language, led many to simply abandon the effort. As one participant succinctly captured this common experience: "If it asks for a login or ID, I often give up halfway" (OA14). Another detailed their failed attempt to reinstall the essential KakaoTalk app: "I tried to reinstall it... but then something popped up with three or four options, and I didn't know what it was. I didn't know which one to choose. So I just closed it" (OA11). These moments represent breakdowns in mental models (\S\ref{sec:mentalmodel})—instances where users' intuitive frameworks for interpreting system behavior fail to provide guidance.

Even when they did persist, learning was an arduous and often frustrating struggle. Many reported needing repeated, step-by-step instructions and frequent refreshers to retain new information. "I need to listen two or three times to barely grasp it... and I forget it all by the next day," one explained (OA06). This sentiment was echoed by another participant who lamented, "I should learn so I don't have to ask my kids, but I forget as soon as I turn around" (OA03). Such accounts reveal how even minor technological updates—like new icons or permission prompts—can create instability for those already managing age-related cognitive or memory challenges.

These learning difficulties also triggered deeper feelings of frustration and inadequacy. For some, the digital world became a source of recurring stress rather than empowerment. Attempts to become more independent were often thwarted not by lack of interest, but by an environment that assumes prior fluency, speed, and linguistic competence. In this context, many developed a personal threshold for what was "worth learning," consciously choosing to stick with functions that had clear, immediate benefits while actively avoiding those that felt unnecessarily complex or unstable. This form of strategic non-learning reflects not a deficit or resignation, but a calculated act of self-preservation—consistent with episodic use patterns (\S\ref{sec:mentalmodel}) where users pause engagement to protect well-being. As one participant stated when asked why she doesn't learn more features: "It's a headache. My daughter tells me to learn, but I don't want to. It's just a headache to learn it" (OA04).

Digital participation thus becomes a delicate balancing act: trying to stay engaged while managing emotional exhaustion and cognitive overload, prioritizing peace of mind over complete digital fluency while also seeking to reduce the burden of constant assistance requests on family members.

\subsubsection{The System-Level Language Barrier as a Potent Trigger for Collaboration}
Language emerged as a persistent, structural barrier that consistently forced participants into collaborative, interdependent problem-solving. While many applications offer Korean language options, the true challenge lies in unavoidable, system-level interactions deeply embedded in the digital ecosystem—including initial device setup, operating system updates, cryptic error messages, privacy prompts, and app store navigation—all of which often default to English and assume high degrees of both linguistic and digital literacy. This pattern of linguistic exclusion forcing reliance on others aligns with scholarship on language barriers in immigrant technology use (\S\ref{nonuse}).

Participants frequently described discomfort when seeking help at U.S. phone stores, where technical explanations were delivered in English at a pace and complexity that left them disoriented. One participant noted, "If I don't understand a phrase, I just say 'Okay,' but I still don't know what to do later" (OA14). Rather than gain clarity, these encounters often resulted in misunderstandings or incomplete setups, requiring repeated visits or outside help. Another shared, "It would be easier if everything were in Korean... I usually have to ask someone younger nearby" (OA09).

These linguistic obstacles become particularly acute and stressful when problems arise, transforming simple technical issues into major, multi-day life disruptions. One participant described their experience when trying to resolve an online banking issue while traveling: "My account got locked..... It took three days to resolve the lock. And you have to wait on the phone for 30 minutes for each call... Then, to do the identity verification, they ask you all these questions like, 'What is your mother's middle name?' or 'What was the color of your previous car?' I couldn't remember what I had entered when I first signed up... It was really, really hard" (OA05).

This example illustrates how seemingly simple digital tasks in the U.S. context are predicated on linguistic and cultural fluency that many older immigrants may not have developed. These linguistic obstacles are not minor inconveniences; they are formidable walls that define the boundaries of digital participation, reinforcing deep-seated dependence on others and reflecting design assumptions about user fluency that often go unexamined. As a result, participants are forced into a constant economy of workarounds: relying on family members as informal translators, memorizing the physical location of buttons instead of reading on-screen instructions, or avoiding entire categories of digital services altogether for fear of making irreversible and unfixable mistakes. In effect, language becomes both a technical and cultural filter, determining what users feel confident doing and what they avoid, making digital tools' promises of accessibility and empowerment fall short when the most basic layers—menus, permissions, instructions—remain linguistically and culturally inaccessible.

\paragraph{Deep Reliance on Others and the "Relational Cost" of Seeking Help}
Faced with these cognitive and linguistic barriers, participants commonly turned to an informal, ad-hoc support network composed of their adult children, grandchildren, community members, or, in some cases, paid aides or retail employees. These help-seeking practices are the clearest and most critical manifestation of interdependent navigation—a model where digital literacy is not an individual attribute but a distributed, shared resource, where the family unit or social circle, rather than the individual, often functions as the digitally proficient actor (\S\ref{nonuse}).

The most common and immediate source of help was adult children. Participants frequently described calling their children for help, often sending photos or videos of frozen screens or unclear error messages to get instructions remotely. This method—part visual cue, part emotional appeal—allowed for real-time problem-solving even across distances. However, this reliance was consistently described as fraught with emotional complexity, reflecting relational uncertainty—the concern not about whether a problem can be solved, but at what social cost. While children were typically portrayed as willing to assist, the act of asking for help—especially repeated requests for the same issue—often led to feelings of tension, embarrassment, or guilt. One participant admitted, "If I ask again, it annoys them, so I end up not using it at all" (OA06). Another described the emotional sting of their child's frustration in a way that revealed deeper fears about aging and dependency:
"My son will get frustrated with me sometimes and say, 'Mom, I already taught you that!'... Another time, he just said, 'If you can't do it, don't use it.' I was so hurt by that. I thought to myself, 'Oh, so when I get older and truly need help, he's going to look down on me and treat me like this.'" (OA15)

These accounts reveal the high "relational cost" associated with interdependence, where the fear of being a burden, appearing incompetent, or straining precious family ties becomes a direct and powerful barrier to technology use and learning—transforming non-use into a relational strategy for preserving family harmony (\S\ref{nonuse}). To avoid this emotional friction, some participants developed sophisticated alternative strategies. One participant explained why she stopped asking her son for help and instead sought commercial assistance:
"I don't want to get the 'look' from my son that I'm a bother, you know? So I'd rather just go to the Verizon store... I go there, and if they help me, I give them a \$20 or \$30 tip. It's more comfortable for me to pay and get it done. Then I don't have to hear any nagging." (OA13)

This demonstrates a shrewd form of navigating interdependence by shifting reliance from a familial relationship, with its complex emotional dynamics, to a transactional one, thereby mitigating the personal cost. In McGranahan's \cite{mcgranahan2016theorizing} terms, this is refusal at its most generative (\S\ref{nonuse}): the rejection of one form of engagement — emotionally costly family help-seeking — produces something new, a commercial support relationship that preserves both dignity and family harmony. OA13 does not simply avoid a situation; she actively constructs an alternative infrastructure for getting things done. 

Beyond family, participants sought help from various community sources, including younger acquaintances at their church or temple, staff at their day-care centers, Medicaid-funded home care aides, and formal classes offered at local Korean community centers (KCC). However, these resources were often described as inconsistent or ultimately ineffective. While smartphone struggles were a common topic of conversation at senior day care centers, "we rarely manage to fix issues properly because we're all of similar age" (OA13). One participant detailed their frustrating experience with a community tech class:
"I attended the computer class at the KCC for six months, but honestly, it was useless. They give you way too much information at once... You learn all these things, but then you go home and you can't practice it, and a day later, you've forgotten everything. You look at your notes, and you don't even know what they mean." (OA13)

These collective accounts paint a clear picture of a critical gap in formal, culturally attuned, and pedagogically sound digital support systems. In the absence of such structures, older adults are left to navigate complex, rapidly changing, and often unforgiving technologies within a fragile and emotionally charged patchwork of informal assistance. What emerges is a form of conditional digital participation—enabled or suspended based on the availability and patience of others—that not only limits independence but also shapes what participants feel is safe or acceptable to explore in the digital world.

\paragraph{Deciding "Not to Learn" as a Protective Choice}
While some older Korean immigrants expressed a desire to improve their digital skills, others described a conscious decision not to engage with certain technologies. This was not necessarily a matter of ability or access, but a protective strategy—a way to manage stress, preserve autonomy, and reduce the risks associated with unfamiliar digital systems. This pattern exemplifies pragmatic disengagement (\S\ref{nonuse}), where refusal represents deliberate strategy rather than failure.

Participants who chose not to adopt new features or apps often did so after weighing the perceived benefits against the cognitive and emotional burden of learning. For instance, some declined to switch from basic phones to smartphones, believing that their communication needs were already met through voice calls and text messaging. As one participant explained, "I only use calls and texts, and I assume my children will help if any issues arise" (OA09). This attitude reflects a form of intentional simplicity—an effort to avoid complexity that might introduce confusion or dependency.

Importantly, this resistance was not rooted in technophobia but revealed a rational risk calculus where participants weighed effort against potential downsides. They often framed their disengagement as a form of self-care—shielding themselves from the emotional toll of trial-and-error learning, the embarrassment of repeated failures, or the fear of financial vulnerability. This connects to episodic use patterns (\S\ref{sec:mentalmodel}), which frame temporal cycles of engagement and abandonment as regulatory mechanisms protecting well-being rather than adoption failures. This selective non-use underscores that digital inclusion must respect individual boundaries, histories, and values rather than simply providing access or instruction. For many older adults, choosing not to engage with certain technologies was a way to retain control over their digital lives rather than surrender to systems that felt alienating or opaque.

\subsection{Evolving Perspectives on Media, Politics, and Emerging Technologies}
\label{findings:evolving}
Beyond practical use, older Korean immigrants also reflected critically on the broader digital media environment, offering nuanced perspectives on algorithmic content, political discourse, and emerging technologies such as AI. Their reflections reveal how emotional well-being, cultural values, and aging shape their willingness to engage with—or retreat from—certain digital spaces and narratives.

As we demonstrate below, participants' responses to algorithmic systems and AI reveal sophisticated folk theories (\S\ref{sec:mentalmodel}) that accurately capture platform dynamics, even as cultural values and uncertainty shape decisions toward strategic non-use (\S\ref{nonuse}). These patterns challenge deficit-oriented assumptions about older adults' digital capabilities while highlighting how rational risk assessment—not technophobia—drives disengagement.

\subsubsection{Intentional Non-Use as a Deliberate Risk Management Strategy}
This practice of pragmatic disengagement was most pronounced and consistently articulated when it came to both financial technologies and politically charged media content. In both domains, participants' resistance was not rooted in vague technophobia but in clear-eyed, rational calculations of risk versus reward, prioritizing emotional well-being and personal security over complete digital engagement. These patterns exemplify "lagging resistance"—deliberate disengagement driven by privacy concerns, emotional fatigue, or cultural discomfort (\S\ref{nonuse}).

\paragraph{Financial Technologies} Online banking, mobile payment apps, and other digital financial services were widely seen as too risky to be worth the convenience—representing a rejection of utilitarian gratification (\S\ref{UGTURT}) when the costs of potential failure outweigh perceived benefits. Participants understood the potential benefits but weighed them against the potential for catastrophic, unmanageable consequences. One participant articulated this risk calculus with perfect clarity:
\begin{quote}
"I don't do online banking. It's not because I don't trust the technology itself, but because I know that if something goes wrong, I don't have the ability to solve it myself. Here in the U.S., the language barrier is a big part of that... I don't use it because I know I wouldn't be able to resolve a problem quickly or effectively." (OA09)    
\end{quote}

Another participant echoed this sentiment, emphasizing the feeling of security that comes with in-person interactions: "I prefer going to the bank... it just feels safer" (OA14). This decision to stick with familiar, analog methods was a deliberate choice to maintain control and avoid situations where they might be rendered helpless by language barriers during high-stakes financial crises. This pattern is in line with refusal as conceptualized by McGranahan \cite{mcgranahan2016theorizing} (see \S\ref{nonuse} for details) than as mere resistance or failed adoption. OA09 does not simply push back against online banking from a position of weakness; rather, she declines to participate on the system's terms and asserts an alternative set of conditions — in-person banking — through which she can engage as an equal. The language barrier here is not an obstacle to be overcome but a legitimate ground for refusal: a culturally coherent repositioning of the terms of digital engagement. 

The same logic was applied to newer digital payment systems like Apple Pay. One participant expressed cautious interest but was ultimately held back by the same fear of unmanageable problems: "I actually want to try using Apple Pay, but I just don't have the courage. I'm worried about what to do if a problem arises and how I would resolve it... It just seems like it would be more of a hassle than it's worth" (OA14).

\paragraph{Political Media Content} Many participants described YouTube as their primary window into Korean political news, allowing them to stay connected with national events and debates. However, this access came at a significant emotional cost—the hedonic cost exceeded the value gained (\S\ref{UGTURT}). The political content was often described as combative, emotionally exhausting, or "noisy"—referring not to loud voices, but to social discord, misinformation, and incivility.

Participants recounted how conservative and liberal channels alike engaged in constant bickering and accusations. One participant noted, "There's so much bickering on both sides that it gives me a headache" (OA04), while another added, "Both sides just hurl insults… I turn it off quickly" (OA04). With conflicting viewpoints and sensational headlines dominating feeds, they often felt unsure of what to believe: "They are so different that you end up not knowing what's right… so I only watch a little and turn it off" (OA08).

\paragraph{Strategic Disengagement as Self-Care}
In both contexts, choosing not to engage was a form of self-care and personal risk management—a way to shield themselves from the emotional toll of trial-and-error learning with finances, the profound fear of financial vulnerability, the potential embarrassment of needing help with sensitive information, and the stress and confusion of polarizing media environments they were already managing in other areas of life. For these participants, the perceived convenience offered by advanced digital technologies and comprehensive media engagement simply did not outweigh the potential for high-stakes, stressful, and seemingly unmanageable problems. Their responses reflect a pragmatic approach that includes not just technical skill, but also the emotional discernment to navigate overwhelming digital environments, prioritizing personal well-being over exhaustive engagement.

\subsubsection{Avoiding the "Noise": Disengaging from Overwhelming and Untrustworthy Media}
The strategy of pragmatic disengagement profoundly shaped participants' media consumption habits, particularly on platforms like YouTube. Many found political content to be combative, polarizing, and emotionally exhausting, frequently using the Korean equivalent of "noisy" not to describe literal volume, but to refer to a broader sense of social discord, misinformation, and incivility. One participant stated bluntly, "There's just so much bickering on both sides that it gives me a headache" (OA02), while another added, "Both sides just hurl insults at each other... I see it and I just turn it off quickly" (OA02). Their retreat from this type of political media was not born from apathy but was an active coping strategy to regulate emotional well-being and preserve peace of mind.

This disengagement also applied to other overwhelming or untrustworthy content categories, such as health information. Participants often encountered contradictory and confusing advice, leading to mistrust and fatigue. One participant described this frustrating dilemma: "You see, for example, with something like grape seed oil. One video might say, 'don't take it, it's bad for you,' while another one says 'you should absolutely take it'—it gets so confusing that I've learned to just turn it off if I feel something is off or exaggerated" (OA06).

Contrary to assumptions about older adults as passive digital users, many participants demonstrated keen awareness of how algorithmic systems shaped their YouTube experiences—revealing accurate mental models of recommendation systems (\S\ref{sec:mentalmodel}). Participants frequently recognized that algorithms promote content with high view counts or strong political leanings, often at the expense of balance or accuracy. One participant remarked, "If you keep watching videos that criticize the government, you might only get similar, biased recommendations" (OA05), reflecting an intuitive grasp of how feedback loops create echo chambers.

Rather than passively accepting algorithmic offerings, many deliberately circumvented them through strategies including manually searching for specific videos, using Korean-language search terms, and avoiding sensational thumbnails—clear instances of selective appropriation (\S\ref{sec:mentalmodel}). As one participant explained, "Since videos with high view counts are shown first, you might only see sensational content. That's why I deliberately try different searches" (OA07).

%Rather than blindly following presented advice, participants developed personal heuristics for evaluating information—cross-checking claims with prior knowledge, evaluating speakers' tone and delivery, or simply disengaging when content felt manipulative or untrustworthy. These practices reflect more than defensive posturing; they reveal algorithmic resistance rooted in self-taught media literacy, where participants understood that viewing behaviors shaped future recommendations and made intentional choices to counteract this influence.

Rather than blindly following presented advice, participants developed personal heuristics for evaluating information—cross-checking claims with prior knowledge, evaluating speakers’ tone and delivery, or simply disengaging when content felt manipulative or untrustworthy. These practices reflect more than defensive posturing; they reveal algorithmic sovereignty rooted in self-taught media literacy, where participants understood that viewing behaviors shaped future recommendations and made intentional choices to counteract this influence. Unlike resistance — which presumes a subordinate actor pushing back against a dominant system — this sovereignty operates by refusing the terms of algorithmic engagement altogether and generating a parallel, community-grounded system of information evaluation (see \S\ref{nonuse}). 

In this context, disengagement was not passive surrender but an active, critical form of evaluation and necessary self-protection strategy in a polluted information environment. By actively filtering, bypassing, or ignoring algorithmic cues, participants asserted agency over their media consumption, prioritizing emotional stability, cultural resonance, and perceived credibility over popularity or entertainment value—demonstrating an emergent form of digital autonomy shaped by life experience, caution, and the pursuit of reliable information on their own terms.

\subsubsection{Media Literacy and Information Discernment}
Participants' digital media use revealed a pragmatic and deeply personal approach to media literacy—one less concerned with mastering formal terminology and more focused on navigating uncertainty, credibility, and overload. While few used the term media literacy itself, their actions reflected a cautious, evaluative mindset: one grounded in intuition, lived experience, and the imperative to protect their well-being.

Health-related content on YouTube was a key site of tension. Participants often encountered contradictory information about supplements, diets, or medical advice, leading to confusion and mistrust. One noted, "One video might say 'don't take grape seed oil' while another says 'you should take it'—it gets so confusing that I just turn it off if I feel something is off" (OA06). Rather than blindly follow advice, many developed personal heuristics—cross-checking information with prior knowledge, evaluating tone and delivery, or simply disengaging when the content felt exaggerated or manipulative.

Participants also connected this need for discernment to broader risks like scams, phishing, or disinformation. Several expressed a sense of responsibility to "study" or stay vigilant, not only for their own safety but to avoid being misled. As one participant put it, "These days, if you don't study media literacy, you can easily be swayed" (OA01). While the term study may not reference structured learning, it signaled a commitment to staying informed and protecting oneself from digital exploitation.

Rather than passive consumption, participants engaged in active filtering—rejecting content that felt untrustworthy, seeking out alternative sources, or consulting with family members when in doubt. This behavior reflects a form of self-directed media stewardship, where credibility is assessed not through credentials or institutions, but through a felt sense of coherence, moral tone, and practical utility.

In this context, media literacy is not a fixed skillset but a survival strategy—one that blends skepticism, cultural attunement, and emotional discernment. For older Korean immigrants, developing these habits was not just about staying informed, but about safeguarding mental clarity.

\subsubsection{A Cautious Curiosity: Attitudes Toward AI and the Technologies of the Future} 
When conversations turned to emerging technologies like Artificial Intelligence (AI), participants expressed a complex blend of cautious curiosity, deep-seated skepticism, and culturally specific concerns—revealing how culturally grounded folk theories (\S\ref{sec:mentalmodel}) profoundly shape technology acceptance. Some recognized potential benefits of AI-enabled tools like ChatGPT or voice assistants for practical tasks such as setting medication reminders, help composing messages, or controlling smart home devices. These conveniences were especially appealing for those managing health conditions or cognitive fatigue. However, this interest was consistently tempered by strong emphasis on maintaining cognitive independence and traditional human skills.

Several participants voiced generational concerns that overreliance on AI could lead to a decline in fundamental human abilities. As one participant thoughtfully noted: "If we start to rely too much on it, humans will become foolish... our brains won't develop. Kids, especially, need to learn basic arithmetic and how to think on their own first" (OA13). Another acknowledged the tension between convenience and skill preservation: "They say it has good features, but I feel it's better to write things myself" (OA16).

Cultural values and spiritual beliefs also significantly shaped their perspectives on AI—illustrating how non-use operates as a values-driven choice (\S\ref{nonuse}) rather than technological limitation. Some participants recounted stories about robotic dolls being used as companions for isolated elderly individuals in South Korea—a concept they found deeply unsettling or spiritually inappropriate. One participant recalled a particularly striking anecdote: "I heard that some elders were so scared of these robotic pets that they would cover them with a blanket at night, out of a fear that they might 'have a soul' or be possessed" (OA20).

These reactions reveal that acceptance of AI, particularly in anthropomorphic forms, is profoundly mediated by deeply rooted cultural and spiritual beliefs about personhood, consciousness, and the sanctity of human connection.

At the same time, participants were not universally opposed to innovation. Practical tools—like puzzle games on large touchscreens, voice-controlled reminders, or interactive digital photo frames—were seen as potentially helpful, especially for supporting cognitive health and aging in place. The key condition for acceptance was meaningful human support: someone to patiently explain the purpose of the tool, demonstrate how to use it, and be available to troubleshoot over time.

Ultimately, their cautious stance reflects not a fear of technology itself, but a profound desire for respectful and meaningful integration—one where digital tools are designed to augment, rather than displace, their most valued human relationships and cultural practices.

\subsubsection{Technology in Dementia Prevention and Cognitive Health}
For older Korean immigrants, the potential of technology to support cognitive health—particularly in the context of dementia or memory decline—elicited cautious optimism. Participants recognized that digital tools could serve as valuable aids for daily routines, memory stimulation, and mental engagement, especially if introduced early and used consistently. However, they also emphasized that the success of such technologies depends on timely training, cultural relevance, and ongoing human support—highlighting the tension between utilitarian promise and the uncertainty that accompanies complex systems (\S\ref{UGTURT}).

Many expressed the belief that digital familiarity needs to be developed before the onset of cognitive decline. One participant shared, "They say if you learn technology before you develop dementia, it will help later" (OA01). This forward-looking logic reflected both a fear of future dependency and a hope that early digital literacy could serve as a buffer—helping them retain autonomy even as memory lapses or confusion increase.

Several participants actively practiced this forward-thinking approach through both physical and digital activities. One participant explicitly connected her engagement with line dancing videos on YouTube to dementia prevention: "They say if you learn to dance when you're older, dementia won't come. So I try to spend a lot of my time on activities like that—watching instructional videos and practicing" (OA17). This integration of digital media consumption into cognitive self-care strategies illustrates how participants viewed technology not merely as a communication tool, but as a resource for maintaining mental acuity. 

This forward-thinking approach was often motivated by a fear of future isolation. One participant, reflecting on her decision to learn computers at age 70, explained: "Before it's too late... I felt that if I become less active, I'll get lonely. So I decided that I should learn computers, just for fun, so I can enjoy something at home" (OA12). However, participants also highlighted the inadequacy of existing digital literacy programs. One described her frustrating experience at a local Korean Community Center class: "They give us—older adults—way too much information at once... When I look at my notes at home, I have no idea what any of it means. I went for six months, and it was all for nothing" (OA13). Another participant captured the broader challenge: "As you get older, you start avoiding anything that requires mental effort—unless it's something you immediately need" (OA11). These accounts underscore a critical tension: while participants recognized the value of early digital engagement for cognitive health, the available learning resources often failed to accommodate their pace, language needs, and emotional thresholds. 

Participants identified specific features they found promising: large screens, simplified interfaces, voice commands, and photo-based memory aids. Some imagined tools that could remind them to take medication or help them engage with meaningful content—such as puzzles using old family photos. These examples reveal a preference for emotionally resonant and intuitive tools over abstract or high-tech solutions.

Yet even with interest in these tools, participants were acutely aware of the barriers. The steep learning curve, lack of Korean-language support, and absence of dedicated tech guidance meant that many of these solutions remained out of reach. One participant reflected on the consequences: "Some people, who have no children and don't know English, end up leaving their devices broken for months until someone they know comes to fix them" (OA03).

This scenario underscores a critical point: access to technology does not guarantee its usefulness. When support systems are missing or inconsistent, even well-designed tools become symbolic rather than functional. Without someone to provide patient instruction and culturally competent guidance, participants risk being further isolated—left behind not only by aging, but by the very technologies meant to support them. This reality underscores that adoption ultimately requires interdependent support (\S\ref{nonuse})—the availability of family members, community resources, or formal services to scaffold meaningful engagement.

\section{Discussion}
%In the sections that follow, we unpack the mechanisms of this pragmatic disengagement through three theoretical lenses. We first examine how cultural values shape \textbf{mental models and algorithmic sovereignty} (\S\ref{cultural}), challenging assumptions of passive adoption. Next, we reframe \textbf{uses and gratifications} through the reality of linguistic risk and selective appropriation (\S\ref{selectiveap}). Finally, we explicate the \textbf{layered uncertainty} that necessitates interdependent navigation (\S\ref{layeredunc}). These theoretical insights provide the foundation for the design (\S\ref{designrec}) and policy (\S\ref{policyrec}) recommendations.

In the sections that follow, we unpack the mechanisms of this pragmatic disengagement through three theoretical lenses. We first examine how cultural values \textbf{shape algorithmic sovereignty} (\S\ref{cultural}), reframing what appears as algorithmic resistance as a culturally grounded form of refusal that redistributes the terms of digital engagement.  Next, we \textbf{reframe uses and gratifications} through the reality of linguistic risk and selective appropriation (\S\ref{selectiveap}). Finally, we explicate the \textbf{layered uncertainty} that necessitates interdependent navigation (\S\ref{layeredunc}). These theoretical insights provide the foundation for the design (\S\ref{designrec}) and policy (\S\ref{policyrec}) recommendations. 

\subsection{Theorizing Selective Engagement as Culturally-Grounded Refusal}
Our findings reveal patterns of strategic disengagement that resonate with Zong and Matias's framework of "data refusal from below" \cite{Zong_Matias_24}, which centers the agency of those who resist algorithmic systems and shifts from individual consent toward collective refusal. Our participants enacted similar forms of refusal—avoiding political YouTube content, rejecting online banking despite understanding its benefits, and delegating high-stakes digital tasks to family members. Crucially, however, we interpret these patterns not merely as resistance but as refusal: an active repositioning of the terms of engagement that "professes a relationship between equals" rather than simply pushing back from a subordinate position \cite[P.320]{mcgranahan2016theorizing}. This distinction matters: where resistance leaves the hierarchy intact, refusal reconfigures it — and our participants' selective disengagement consistently produces new relational and informational infrastructures that operate on their own cultural terms. 

Critically, their refusals were filtered through culturally specific logics of risk, dignity, and intergenerational obligation that extend Zong and Matias's framework in important ways.
Where data refusal from below emphasizes collective resistance to datafication and surveillance, our participants' non-use operated through family networks shaped by Korean values around face-saving and filial piety, creating what we term \textit{pragmatic disengagement}—partial, conditional, and mediated participation that is neither fully resistant nor fully compliant but represents culturally-situated wisdom about living well with technology under linguistic marginalization.

This pragmatic disengagement connects to broader literature showing that non-use emerges from social circumstances \cite{light_akama_2012} and emotional costs. What makes our participants' disengagement distinctively culturally specific — rather than a universal user behavior — is the particular moral economy through which it operates. Decisions to disengage are not made by isolated individuals calculating personal risk; they are embedded in Korean cultural values around nunchi (social attunement), chemyeon (face-saving), and filial piety that shape what counts as an acceptable cost, who bears it, and how it is managed. OA13's preference for paying a Verizon employee rather than asking her son reflects not just a practical workaround but a culturally specific calibration of dignity, obligation, and relational harmony that is irreducibly Korean in its logic — and that distinguishes pragmatic disengagement from the generic technology avoidance that any user might practice. For example, participants' avoidance of online banking (OA09) reflected assessments of social and emotional costs, not skill deficits, echoing Sheehan and Le Dantec's "functional minimalism" \cite{sheehan_23}.
The family-centered nature of technology practices extends research on intergenerational mediation \cite{wong-villacres_et_al_19, yuan_et_al_24} by showing how family assistance creates both support and tension—valued yet threatening to dignity and autonomy.
Language emerged as a central axis of non-use, with participants' abandonment of tools when confronted with untranslated prompts paralleling documented patterns of linguistic exclusion in medical portals \cite{le_et_al_24} and voice assistants \cite{mittal_et_al_25}, operating across multiple layers from interfaces to error recovery.

\subsubsection{Culturally-Grounded Mental Models and Algorithmic Sovereignty} \label{cultural}
We interpret our participants' engagement through the lens of "folk theories," where users interpret opaque algorithmic systems through local cultures and moral economies \cite{devito2017algorithms, kelly_et_al_23, bijker1997bicycles}.
When OA20 described covering robotic companions with blankets out of concern they might have souls, she revealed mental models shaped by Buddhist and Korean folk beliefs.
Similarly, OA05's recognition that viewing government-critical videos leads to "biased recommendations" and OA02's deliberate varied searching to avoid sensational content demonstrate sophisticated "folk algorithmic literacy."

%Contrary to assumptions of passivity, these mental models produced active resistance strategies we term \textit{algorithmic sovereignty}.
%OA16's practice of only watching cooking channels that "align with my values" and rejecting content with "too much sugar" shows how personal boundaries override algorithmic suggestion.
%Designers must recognize that these culturally grounded models often produce protective behaviors that serve user interests better than naive engagement; resistance is an active preservation of valued human capacities rather than a failure of adoption \cite{bucher2019algorithmic}.

Contrary to assumptions of passivity, these folk theories produce practices of algorithmic sovereignty that are simultaneously acts of refusal and acts of generation. OA16's practice of only watching cooking channels that "align with my values" and rejecting content with "too much sugar" illustrates how cultural and moral values redefine the terms of platform engagement rather than simply opposing them. This refusal does not end participants' relationship with YouTube; it generates an alternative one. When participants bypass the recommendation feed and instead circulate pre-vetted links through KakaoTalk, discuss content collectively at senior centers, and apply community-developed heuristics for evaluating trustworthiness, they produce a parallel knowledge infrastructure embedded in social relationships and Korean communal values. This is precisely what McGranahan  means when she argues that refusal is generative: "the ending of one thing is often the generation of something new" \cite[P.322]{mcgranahan2016theorizing}. What distinguishes this from generic algorithmic avoidance is its collective, community-grounded character: participants do not navigate platform logics as isolated individuals but as members of a tightly networked Korean diaspora community whose shared heuristics — stick to known channels, avoid the home feed, verify through KakaoTalk — constitute a culturally specific alternative to algorithmic curation. Designers must recognize that these culturally grounded refusals often produce protective behaviors that serve user interests better than naive engagement, and that design solutions must support rather than circumvent these community-generated epistemic practices \cite{bucher2019algorithmic}. 

\subsubsection{Selective Appropriation and Culturally-Filtered Gratifications}
\label{selectiveap}
Our findings complicate Uses and Gratifications Theory \cite{katz1973uses} by demonstrating how cultural distance transforms the calculus of utility and pleasure.
While platforms like KakaoTalk served as a "lifeline" for family bonds (OA04), the linguistic barriers inherent in these tools introduced a \textit{linguistic risk premium}.
OA09's sophisticated cost-benefit reasoning explained that despite understanding efficiency gains, the inability to navigate error recovery in English made the risk of online banking catastrophic. This is not a failure of digital literacy but a refusal as defined in McGranahan \cite{mcgranahan2016theorizing}: OA09 declines to participate in a system whose conditions — English-language error recovery, culturally alien security verification questions — position her as a subordinate user rather than an equal participant. By choosing in-person banking instead, she asserts her own conditions of engagement, redistributing the power relationship between herself and the financial system on terms she can control \cite[P.368]{mcgranahan2018refusal}. The Korean cultural context matters here: the cascading shame and loss of face that would accompany a public financial failure in front of English-speaking bank staff is not a universal concern but a specifically Korean emotional calculus that makes refusal the rational, dignified choice.

Consequently, engagement exemplified "appropriation work" \cite{salovaara2011everyday}, where users repurposed technology to serve specific needs while strictly walling off intrusive features.
OA02's rejection of voice-to-text because "English words get oddly converted" and OA06's practice of turning off videos when content seemed exaggerated illustrate how appropriation is filtered through linguistic and emotional constraints.
This temporal flexibility—ramping engagement up for information and down for mental health—represents a sophisticated form of self-regulation and "partial domestication" that traditional adoption models fail to capture \cite{sorensen2006domestication}.

\subsubsection{Layered Uncertainty and Interdependent Navigation}
\label{layeredunc}
Building on Uncertainty Reduction Theory \cite{kramer1999motivation}, we identify \textit{layered uncertainty} as a barrier where multiple ambiguities compound to paralyze action.
We identify four interconnected planes:
\begin{itemize}
    \item \textbf{Technical uncertainty}: OA11's abandonment of app reinstallation when faced with multiple unexplained options illustrates how technical ambiguity produces paralysis.
    \item \textbf{Linguistic uncertainty}: OA05's three-day banking crisis exemplifies how English-language prompts create compounding barriers—each step representing a potential obstacle.
    \item \textbf{Relational uncertainty}: OA06's admission about not asking again to avoid annoying family revealed how concerns about relationship costs shape technology decisions.
    \item \textbf{Epistemic uncertainty}: Contradictory health information generated protective skepticism that limited willingness to trust new sources.
\end{itemize}

Participants managed this cumulative uncertainty through \textit{interdependent navigation}, where digital literacy functions as a distributed, relational resource.
OA15's exchange of sending a faucet photo to receive a how-to video represents cooperative work distributing problem-solving across generations.
However, this interdependence carries emotional costs; OA13's preference for paying Verizon employees rather than asking her son to avoid "the look" showed how commercial transactions substitute for emotionally costly family support. This substitution exemplifies the generative dimension of refusal \cite{mcgranahan2016theorizing}: OA13 does not merely avoid an uncomfortable situation but actively constructs a new support infrastructure — commercial, transactional, dignity-preserving — that allows her to maintain both technological participation and family harmony. The refusal of emotionally costly help-seeking produces something new, and that something is culturally specific: the willingness to pay a stranger rather than burden a child reflects Korean norms of intergenerational obligation and chemyeon that make family-directed help-seeking emotionally costly in ways that cross-cultural frameworks of interdependence do not fully capture. 

Our findings extend Wong-Villacres et al.'s \cite{wong-villacres_et_al_19} concept of intergenerational mediation by reframing it not just as support, but as critical relational infrastructure necessary for adoption. Traditional digital literacy training often fails because it prioritizes individual skill acquisition. Therefore, successful adoption for this demographic requires the continuous availability of patient, culturally competent human scaffolding. Taken together, the three dimensions of our analysis — algorithmic sovereignty as refusal, linguistically-filtered gratifications as refusal, and interdependent navigation as the relational infrastructure that makes selective engagement sustainable — reveal a consistent underlying logic: our participants refuse digital participation on systems' terms and generate, through that refusal, alternative modes of engagement that are culturally coherent, relationally embedded, and dignified. This is not the story of a community failing to adopt technology; it is the story of a community practicing what McGranahan calls the "insistence on the possible over the probable" — refusing the terms of inclusion that existing systems offer and asserting the right to participate differently \cite[P.323]{mcgranahan2016theorizing}. 

\subsection{Design Recommendations: Toward Culturally-Grounded and Emotionally-Attuned Interfaces}
\label{designrec}
Our theoretical analysis yields specific design implications. If mental models are culturally-grounded, interfaces must accommodate diverse interpretive frameworks rather than assuming universal logics. If gratifications are filtered through relational and linguistic risk, systems must reduce the stakes of failure. If uncertainty is layered and managed through interdependent navigation, tools must support collaborative use while preserving dignity.

We propose three interconnected design directions:

\begin{enumerate}
    \item \textbf{Full-stack language support}: Current approaches often provide translated interfaces while leaving system-level interactions—settings, permissions, error messages, security prompts—in English. Our findings reveal that these untranslated layers create the highest-stakes barriers. Moreover, such linguistic gaps carry normative weight, signaling to communities which users systems are designed for \cite{change_et_al_2024}. True linguistic inclusion requires bilingual support extending through troubleshooting pathways and error recovery, precisely the moments when users are most vulnerable and least able to seek help. This includes culturally-appropriate security questions and verification processes that do not assume Western naming conventions or life experiences.
    
    \item \textbf{Dignity-preserving collaborative features}: Given the reality of interdependent navigation, systems should support family involvement while protecting older adults' autonomy and dignity. Given the reality of interdependent navigation, systems should support family involvement while protecting older adults' autonomy and dignity \cite{druga_et_al_22, stefanidi_et_al_24, silva_et_al_24}. Drawing on our finding that relational uncertainty shapes technology decisions, we propose: (a) \textit{asymmetric visibility} features where helpers can assist without older adults feeling surveilled; (b) \textit{delayed assistance options} allowing users to struggle privately before help becomes visible; and (c) \textit{success acknowledgment} celebrating independent task completion. These features address the emotional labor of help-seeking that current designs ignore.
    
    \item \textbf{Emotion-sensitive content curation}: Participants' sophisticated algorithmic resistance suggests demand for simplified interfaces that spotlight essential tools through large, recognizable icons rooted in existing mental models and daily routines \cite{johnson1989mental, devito2017algorithms}, alongside explicit controls over content exposure. Rather than forcing users to develop workarounds, platforms could offer straightforward options to opt out of algorithmically suggested content, reduce exposure to political material, flag contradictory health information, or prioritize trusted sources \cite{karizat_et_al_21, joy_et_al_25}. This dual approach addresses both interface complexity and content overwhelm, reducing algorithmic anxiety \cite{jhaver_et_al_18} while respecting participants' demonstrated preference for emotional wellbeing over comprehensive engagement.
.
\end{enumerate}

\subsection{Policy Recommendation: Community-Based Digital Mediation as Formal Care}
\label{policyrec}
Building on Kuoppamaki et al.'s \cite{kuoppamaki2022enhancing} work on community-based digital inclusion, our findings reveal the need for more sustained, culturally embedded support. The failures of Korean Community Center classes (OA13) echo Kuoppamaki's concerns about one-size-fits-all training but suggest the solution isn't just better pedagogy—it's recognizing digital mediation as ongoing relational work rather than knowledge transfer. This aligns with Brewer et al.'s \cite{brewer_et_al_21} findings on older adults avoiding online discussions to manage social tensions. Our participants' strategic disengagement from political YouTube content reflects similar boundary-setting, suggesting that successful community programs must respect these protective strategies rather than trying to overcome them. Given our findings, we present the following policy recommendations: 

    \begin{enumerate}
        \item \textbf{Integrate Tech Support into Existing Care Infrastructure}: Train Medicaid-funded home care aides and senior daycare staff to provide basic digital mediation, such as resetting devices or guiding users through app updates. Recognize this support as essential—not incidental—carework, particularly in linguistically isolated households.
        \item \textbf{Support Community Tech Liaisons}: Fund bilingual digital mediators located at trusted sites (e.g., Korean churches, temples, senior centers) who can provide culturally competent, on-call tech assistance. These liaisons would serve as interpreters, educators, and advocates—closing gaps between users and institutions \cite{kuoppamaki2022enhancing}.
        \item \textbf{Design Interest-Based, Culturally Anchored Digital Literacy Modules}: Facilitate peer-led programs centered on culturally resonant content—such as watching Korean-language cooking videos, sharing church livestreams (similar to proposal by Brewer et al. \cite{brewer_et_al_21}), or reading news from Korean media outlets—so that learning is joyful, not burdensome, and socially meaningful.
    \end{enumerate}

By formalizing these practices through public and community partnerships, digital inclusion becomes not just about skills acquisition, but about fostering digital belonging across generations and cultural contexts.

\section{Limitations and Future Work}
Our sample primarily included older Korean immigrants in a specific metropolitan region, which may limit the generalizability of our findings to other immigrant communities or geographic contexts. Additionally, the study relied on voluntary participation, which may have excluded individuals who are more socially isolated or have extremely limited exposure to technology—precisely those who might face the most acute barriers to digital inclusion.

Additionally, our recruitment was concentrated within a Buddhist temple community, which may not fully represent the diversity of religious affiliations among older Korean immigrants. Korean Christian churches constitute a significant portion of Korean American ethnic institutions and often serve as primary social hubs for older adults. Future research should include participants from Christian congregations and secular organizations to examine whether religious community context shapes digital practices and support-seeking behaviors. 

Future research could employ longitudinal or mixed-method designs to explore how digital use, non-use, and partial adoption evolve over time—especially in relation to cognitive changes, caregiving dynamics, or shifting family support networks. A more temporally sensitive approach could reveal how pragmatic disengagement patterns emerge or shift with aging, illness, or life transitions.

Building on our design recommendations, we also suggest future work incorporate co-design workshops to validate and refine culturally aligned technology solutions. A model can be drawn from Liaqat et al. \cite{Liaqat_et_al_21}, who employed participatory storytelling tools with intergenerational family members to surface tensions around privacy, authorship, and digital agency. Similarly, workshops engaging older immigrants, caregivers, and community leaders could help assess the usability, emotional resonance, and ethical fit of proposed features—such as bilingual interfaces, simplified shared-use modes, and emotion-sensitive content curation.

Finally, expanding inquiry into AI-based eldercare technologies—including robotic companions, remote monitoring tools, and personalized health AI assistants—could deepen our understanding of how older immigrants perceive automation in relation to trust, cultural values, and emotional safety. Such work would benefit from attending not only to functional adoption but to the symbolic and affective meanings these technologies carry within diverse aging contexts.

\section{Conclusion}
Older Korean immigrants navigate digital environments not through passive limitation but through deliberate, culturally grounded strategies. We identify two core practices: \textit{pragmatic disengagement}—the selective refusal of emotionally taxing or linguistically risky digital engagement—and \textit{interdependent navigation}—a distributed mode of digital literacy enacted through family, community, and commercial networks rather than individual skill. Together, these practices constitute a form of digital citizenship shaped by Korean cultural values and enacted on participants' own relational terms. Designing for this reality demands that digital inclusion center not access alone, but the language, emotion, culture, and community through which meaningful participation becomes possible.
\section{Acknowledgments}

\bibliographystyle{ACM-Reference-Format}
\bibliography{sample-base}

\end{document}